\documentclass[longauth,oldversion,letter]{aa}

\usepackage{graphicx}
\usepackage{natbib}
\bibpunct{(}{)}{;}{a}{}{,} 

\begin{document}

\newcommand{\Cerenkov}{Cherenkov\ }
\newcommand{\astar}{Sgr~A$^{\star}$}
\newcommand{\HMS}[3]{$#1^{\mathrm{h}}#2^{\mathrm{m}}#3^{\mathrm{s}}$}
\newcommand{\DMS}[3]{$#1^\circ #2' #3''$}
\newcommand{\TODO}[1]{\textbf{TODO: \emph{#1}}}
\renewcommand{\cite}{\citep} 

\title{Simultaneous H.E.S.S. and Chandra observations of Sgr~A$^{\star}$ \\ during an X-ray flare}

\titlerunning{$\gamma$-ray/X-ray observations of Sgr~A$^{\star}$ during a flare}
 
\authorrunning{The H.E.S.S. Collaboration}

\author{
  F. Aharonian\inst{1,13}
 \and A.G.~Akhperjanian \inst{2}
 \and U.~Barres de Almeida \inst{8} \thanks{supported by CAPES Foundation, Ministry of Education of Brazil}
 \and A.R.~Bazer-Bachi \inst{3}
 \and Y.~Becherini  \inst{12}
 \and B.~Behera \inst{14}
 \and W.~Benbow \inst{1}
 \and K.~Bernl\"ohr \inst{1,5}
 \and C.~Boisson \inst{6}
 \and A.~Bochow \inst{1}
% ...
}

\institute{
Max-Planck-Institut f\"ur Kernphysik, P.O. Box 103980, D 69029
Heidelberg, Germany
\and
 Yerevan Physics Institute, 2 Alikhanian Brothers St., 375036 Yerevan,
Armenia
\and
Centre d'Etude Spatiale des Rayonnements, CNRS/UPS, 9 av. du Colonel Roche, BP
4346, F-31029 Toulouse Cedex 4, France
\and
Universit\"at Hamburg, Institut f\"ur Experimentalphysik, Luruper Chaussee
149, D 22761 Hamburg, Germany
\and
Institut f\"ur Physik, Humboldt-Universit\"at zu Berlin, Newtonstr. 15,
D 12489 Berlin, Germany
\and
LUTH, Observatoire de Paris, CNRS, Universit\'e Paris Diderot, 5 Place Jules Janssen, 92190 Meudon, 
France
Obserwatorium Astronomiczne, Uniwersytet Ja
\and
IRFU/DSM/CEA, CE Saclay, F-91191
Gif-sur-Yvette, Cedex, France
\and
University of Durham, Department of Physics, South Road, Durham DH1 3LE,
U.K.
\and
Unit for Space Physics, North-West University, Potchefstroom 2520,
    South Africa
\and
Laboratoire Leprince-Ringuet, Ecole Polytechnique, CNRS/IN2P3,
 F-91128 Palaiseau, France
\and 
Laboratoire d'Annecy-le-Vieux de Physique des Particules, CNRS/IN2P3,
9 Chemin de Bellevue - BP 110 F-74941 Annecy-le-Vieux Cedex, France
\and
Astroparticule et Cosmologie (APC), CNRS, Universite Paris 7 Denis Diderot,
10, rue Alice Domon et Leonie Duquet, F-75205 Paris Cedex 13, France
\thanks{UMR 7164 (CNRS, Universit\'e Paris VII, CEA, Observatoire de Paris)}
\and
Dublin Institute for Advanced Studies, 5 Merrion Square, Dublin 2,
Ireland
\and
Landessternwarte, Universit\"at Heidelberg, K\"onigstuhl, D 69117 Heidelberg, Germany
% ...
}

\offprints{j.a.hinton@leeds.ac.uk}

\abstract{

The rapidly varying ($\sim$10 minute timescale) non-thermal X-ray
emission observed from Sgr A$^{\star}$ implies that particle acceleration
is occuring close to the event horizon of the supermassive black
hole. The TeV $\gamma$-ray source HESS\,J1745$-$290 is coincident with
Sgr A$^{\star}$ and may be closely related to its X-ray
emission. Simultaneous X-ray and TeV observations are required to
elucidate the relationship between these objects. We report on
joint H.E.S.S./Chandra observations performed in July 2005, during which an
X-ray flare was detected.  Despite a factor of $\approx$9 increase in the
X-ray flux of Sgr~A$^{\star}$, no evidence is found for an increase in
the TeV $\gamma$-ray flux from this region. We find that an increase
in the $\gamma$-ray flux of a factor of 2 or greater can be excluded at a
confidence level of 99\%. This finding disfavours scenarios in which
the keV and TeV emission are associated with the same population of
accelerated particles and in which the bulk of the $\gamma$-ray
emission is produced within $\sim$$10^{14}$ cm ($\sim$100$\,R_{S}$) of
the supermassive black hole.  }


\maketitle

\section{Introduction}

Measurements of stellar orbits in the central parsec of our galaxy
have revealed the existence of a supermassive, $(3.6\pm0.3)\times
10^{6}$ solar mass, black hole coincident with the radio source
Sgr~A$^{\star}$~\cite{GC:Eisenhauer05}. The compact nature of
Sgr~A$^{\star}$ has been demonstrated both by direct VLBI
measurements~\cite{GC:Shen05} and by the observation of X-ray and near
IR flares with timescales as short as a few minutes (see for example 
\citet{GC:Porquet08}, \citet{GC:Eckart06} and \citet{GC:Porquet03}).  
Variability on these timescales limits the emission region to within $<10$ Schwarzschild
radii ($R_{S}$) of the black hole.  X-ray flares from \astar\ reach peak luminosities of
$4\times10^{35}$~erg~s$^{-1}$, two orders of magnitude brighter than
the quiescent value~\cite{GC:Porquet03, GC:Baganoff03}, and exhibit a
range of spectral shapes~\cite{GC:Porquet08}.  Several models
of the origin of this variable emission exist, many of which invoke
non-thermal processes close to the event horizon of the central black
hole to produce a population of relativistic particles
\citep[see e.g.][]{GC:Markoff01,GC:Yuan03,GC:AharonianNeronov05apj,
GC:Liu06,GC:Liu06HESS}.

Model-independent evidence that ultra-relativistic
particles exist close to Sgr~A$^{\star}$ can be provided by the observation
of TeV $\gamma$-rays from this source. Indeed, TeV $\gamma$-ray
emission has been detected from the Sgr~A region by several
ground-based instruments~\cite{GC:Whipple04,GC:CANGAROO,HESS:gc04,
GC:MAGIC06}. The most precise measurements of this source,
HESS\,J1745$-$290, are those performed using the H.E.S.S. telescope
array. The centroid of the source is located $7'' \pm
14_{\mathrm{stat}}'' \pm 28_{\mathrm{sys}}''$ from Sgr~A$^{\star}$,
and has an rms extension of $<1.2'$~\cite{HESS:gcprl}, with work
underway to reduce these uncertainties~\cite{GC:ICRCHESSPos}.  

TeV emission from \astar\ is expected in several models of particle
acceleration in the environment of the black hole.  In some of these
scenarios \cite{GC:Levinson, GC:AharonianNeronov05apj}, TeV emission is produced in the
immediate vicinity of the SMBH, and variability is expected. In
alternative scenarios, particles are accelerated close to \astar\ but radiate
within the central $\sim$10~parsec region~\cite{GC:AharonianNeronov05}, or are
accelerated at the termination shock of a wind driven by the SMBH
\cite{GC:AtoyanDermer}.  However, several additional candidate objects exist for
the origin of the observed $\gamma$-ray emission. The radio centroid
of the supernova remnant (SNR) Sgr~A East lies $\sim 1'$ from
Sgr~A$^{\star}$, only marginally inconsistent with the position of the
TeV source presented by \citet{HESS:gcprl}. Shell-type SNR are now well
established TeV $\gamma$-ray sources~\cite{HESS:rxj1713p3,HESS:velajnr2}
and several authors have suggested that Sgr A East is the origin of the TeV
emission \citep[see for example][]{GC:Crocker}. 
However, improvements in the uncertainty in the centroid position of
HESS\,J1745$-$290 \cite{GC:ICRCHESSPos} effectively exclude the possibility of Sgr~A East being the dominant
$\gamma$-ray source in the region.  The 
pulsar wind nebula candidate G\,359.95$-$0.04 discovered by \citet{GC:Wang06} is located only 9$''$ from \astar\ and can
plausibly account for the TeV emission~\cite{GC:Hinton07}. Particle
acceleration at stellar wind collision shocks within the central young
stellar cluster has also been hypothesised to explain the $\gamma$-ray
source~\cite{GC:Quataert05}. Finally, the possible origin of this source in the
annihilation of WIMPs in a central dark matter cusp has been
discussed extensively~\cite{GC:Hooper04,GC:Profumo05,HESS:gcprl}.

Given the limited angular resolution of current VHE $\gamma$-ray
telescopes, the most promising tool in identifying the TeV
source is the detection of \emph{correlated variability} between the
$\gamma$-ray and X-ray, and/or NIR regimes.  A significant increase in
the flux of HESS\,J1745$-$290, occuring simultaneously with a flare 
detected in a waveband with sufficient angular resolution to isolate \astar, would unambiguously identify the $\gamma$-ray source.  Therefore,
whilst not all models of the TeV emission from Sgr~A$^{\star}$ predict
variability in the VHE source, coordinated IR/keV/TeV observations can
be seen as a key aspect of the ongoing program to understand the
nature of this enigmatic source.

\section{Observations and Results}

A coordinated multiwavelength observing campaign targeting \astar\ 
was performed during July/August 2005. As part of this campaign, observations with
H.E.S.S. occurred for 4-5 hours each night from July 27 to August 1
(MJD 53578-53584). Four Chandra observations with IDs 5950-5954 took
place between July 2 and August 2.  A search for flaring events in
the X-ray data yielded two significant events during the Chandra
campaign, the first during observation (obs.) ID 5952 on July 29, and the
second during obs. ID 5953 on July 30.  The second of
these flares occurred during a period of H.E.S.S. coverage and is
described in detail here.

\begin{figure}[ht]
  \centering
  \resizebox{1.01\hsize}{!}{\includegraphics{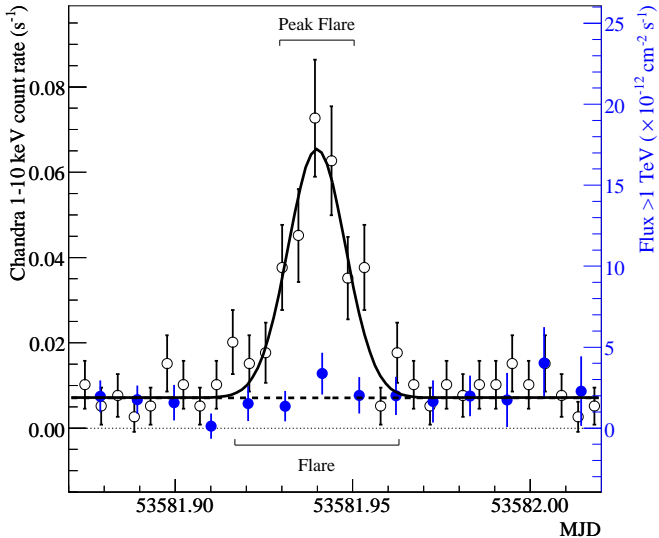}} 
    \caption{ X-ray and $\gamma$-ray light curves for the Galactic
      Centre on MJD 53581.  The open circles show the (background-subtracted)
      Chandra 0.3--8 keV count rate from within $2.5''$ of \astar\ in 400-second bins. The X-ray
      flare is well described by a Gaussian (solid curve), and the time
      periods labelled \emph{Flare} and \emph{Peak Flare} are those used 
      for the X-ray spectral analysis. The closed circles show the 
      VHE $\gamma$-ray light curve from H.E.S.S. in 15 minute bins,
      scaled such that the historical VHE flux level \cite{HESS:gcprl}
      (dashed line) matches the quiescent X-ray count-rate.
    }
  \label{fig:1}
\end{figure}

The 49~ks of ACIS-I data from obs. ID 5953 were analysed using CIAO
version 3.4 and a light curve was extracted from a circular aperture of
radius 2.5$''$, centred on \astar\,(RA 17$^{\rm h}$45$^{\rm m}$40.039$^{\rm s}$,
Dec. -29$^{\circ}$00$'$28.12$''$). Consistent results were obtained using 
a 1.5$''$ aperture. The background level was estimated from a
surrounding region of 8.3$''$ radius, offset by 5.8$''$ to the East of
\astar\ to avoid contamination from the stellar complex IRS\,13 \citep{GC:IRS13} 
and G\,359.95$-$0.04. All
photons in the energy range of 300~eV to 8~keV were included in the
analysis.  The resulting background-subtracted light curve (with 400~s binning)
is shown in Fig. \ref{fig:1}. A significant flare, peaking at MJD
$53581.940\pm0.001$, was detected.  Before and after the flare the event
rate was consistent with a constant value of ($7.1\pm0.1$) counts
ks$^{-1}$, which is consistent with the level found by \citet{GC:Baganoff03}.
The shape of the flare is well described by a Gaussian of full width
half maximum $t_{\rm flare}=$(1.6$\pm$0.2)~ks. No indication of
additional variation or significant substructure was found when
testing residuals with 200~s, 500~s, and 1500~s binning.  The flare
reached a peak level of ($65 \pm 9$) counts ks$^{-1}$ (from the
Gaussian fit), $\approx9$ times the quiescent level.  The flare
duration is comparable with that of other flares detected previously
from \astar\ \citep[for example][]{GC:Eckart06}, and is amongst the
brightest detected by Chandra so far with a net
integrated signal of ($101\pm11$) counts.

The $\gamma$-ray data consist of 72 twenty-eight minute runs, 66 of
which pass all quality selection cuts described by
\citet{HESS:crab}. All runs on the night of the X-ray flare pass these
cuts and we find no evidence for cloud cover in 
simultaneous sky temperature (radiometer) measurements
\citep[see][]{HESS:crab,HESS:atmosphere}.  The data were analysed using the
H.E.S.S. standard \emph{Hillas parameter} based method with the
\emph{standard} $\gamma$-ray selection cuts 
(including a cut on angular distance from 
\astar\ of 6.7$'$) described in
\cite{HESS:crab}, resulting in an energy threshold of 160~GeV. 
There is no evidence of variations in the flux on
timescales of days, and the mean $\gamma$-ray flux $F(>1\,\mathrm{TeV})$ for this week
of observations was ($2.03 \pm 0.09_{\mathrm{stat}} $)$\times 10^{-12}$
cm$^{-2}$ s$^{-1}$, consistent with the average value for
H.E.S.S. observations in 2004, ($1.87 \pm 0.1_{\mathrm{stat}} \pm
0.3_{\mathrm{sys}}$)$\times 10^{-12}$ cm$^{-2}$
s$^{-1}$~\cite{HESS:gcprl}. An independent analysis based on the \emph{Model
Analysis} method described by \citet{deNaurois:Model}, produced
consistent results. 

The time window for the $\gamma$-ray analysis is defined to be the
region within $\pm1.3\sigma$ of the best-fit peak time of the X-ray
flare (containing $\approx80$\% of the signal).  The mean flux within
this window (marked ``Flare'' in Figure~\ref{fig:1}) is
$F(>1\,\mathrm{TeV})\,=$($2.05\pm 0.76$)$\times\,10^{-12}$ cm$^{-2}$
s$^{-1}$. This flux level is almost identical to the mean flux level
for the entire week of observations.  There is, therefore, no evidence
for an increase in $\gamma$-ray flux of HESS\,J1745$-$290 during the
X-ray flare and a limit to the relative flux increase of less than a
factor 2 is derived at the 99\% confidence level. In principle, a
(positive or negative) time lag might be expected between the X-ray
flare and any associated $\gamma$-ray flare. The existence of a
counterpart $\gamma$-flare with a flux increase of a factor $\gg2$
(relative to the mean $\gamma$-ray flux level) requires a lag of at
least 80 minutes (to fall outside the period of
H.E.S.S. observations).

The results of a spectral analysis of the X-ray emission from \astar\
during obs. ID 5953 are presented in Table~\ref{tab}. Spectra are
given for the entire period (\emph{Overall}), intervals of $\pm2$~ks
(\emph{Flare}) and $\pm0.9$ ks (\emph{Peak Flare}, $\pm1.3\sigma$)
around the maximum, and for the part of the dataset outside $\pm3$~ks
of the maximum (\emph{Quiescent}).
For all four data-sets, the background subtracted spectra were fitted
with an absorbed  power-law model using a
fixed value of $n_{\rm H}=9.8 \times 10^{22}$ cm$^{-2}$, as found
from fitting the (\emph{Overall}) dataset with  $n_{\rm H}$ free, which is consistent 
with the value found by \citet{GC:Baganoff03} of $(9.8^{+4.4}_{-3.0})\times 10^{22}$ cm$^{-2}$.
The flare spectrum
is harder than that found in the quiescent state (at the $2.5 \sigma$ level).
The quiescent state spectrum is consistent
with that found previously \cite{GC:Baganoff03}.

\begin{table}[t]
\begin{center}
\begin{tabular}{|l|c|c|c|} \hline
                 & $\Gamma$       &   $F_{2-10}$  &    $\chi^{2}$/dof  \\ \hline
\tiny		 & 		  &  		&      	     \\
\emph{Overall}   & 2.58$\pm$0.20  &   2.71     &    24.6/24   \\
		 & 		  &  		&      	     \\
\emph{Quiescent} & 3.08$\pm$0.27  &   1.82     &    16.3/17   \\
		 & 		  &  		&    	     \\
\emph{Flare}     & 1.83$\pm$0.43  &   14.8     &    2.88/6    \\
	\tiny	 & 		  &  		&    	     \\
\emph{Peak Flare}& 1.70$\pm$0.41  &   19.0    &    3.1/5    \\\hline
\end{tabular} 
\end{center}
\caption{
Parameters of absorbed power-law ($dN/dE\propto E^{-\Gamma}$) fits to the 
0.2--10 keV spectral data of \emph{Chandra} obs. 5953.
$F_{2-10}$ is the absorbed model flux between 2 and 10~keV in units of  $10^{-13}$ erg cm$^{-2}$ s$^{-1}$.
The absorbing column $n_{H}$ is fixed at $9.8\times 10^{22}$ cm$^{-2}$ for all fits. 
Uncertainties are expressed as 1 standard deviation errors.
}
\label{tab}
\end{table}

\begin{figure*}
  \centering
  \resizebox{0.8\hsize}{!}{\includegraphics{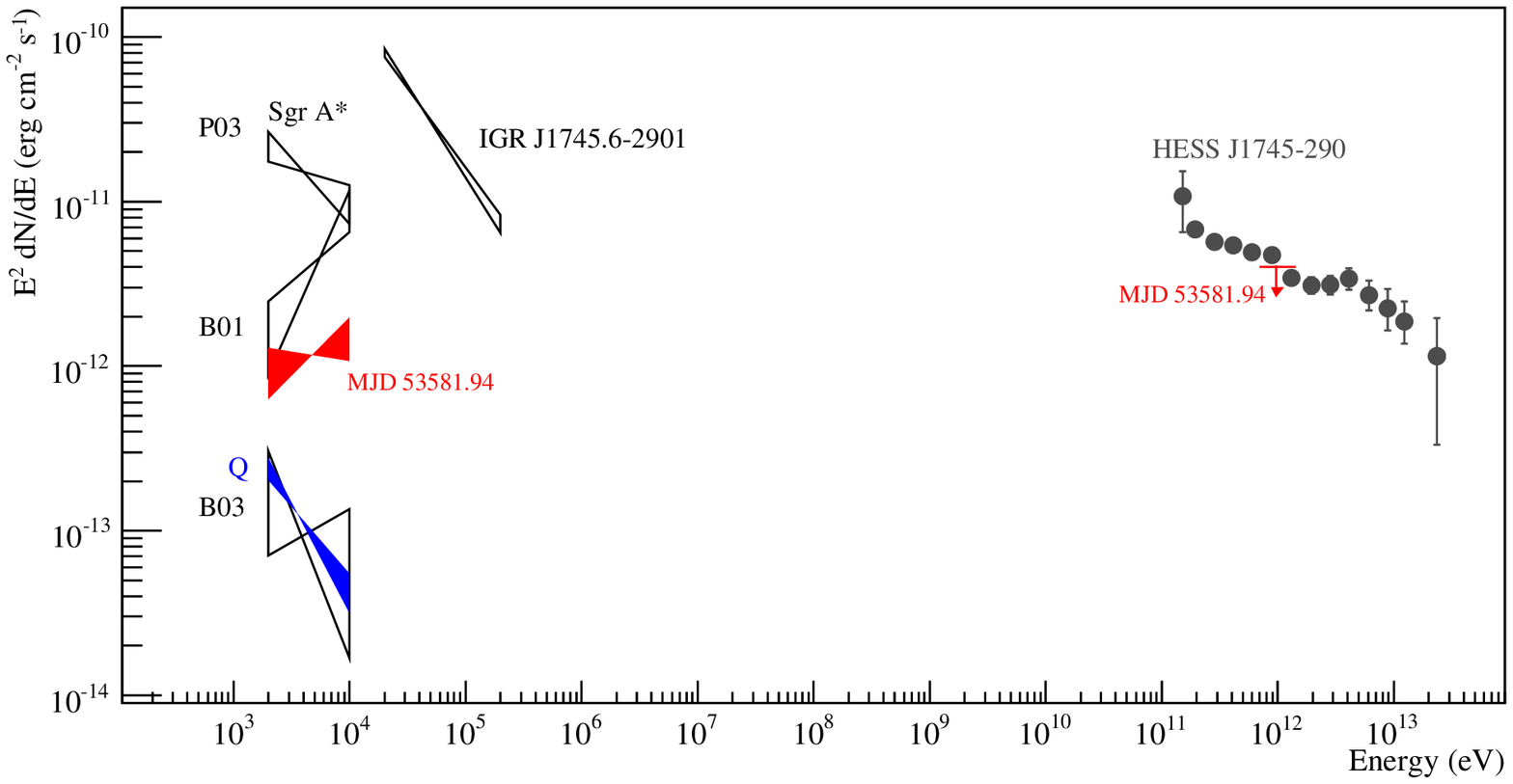}} 
    \caption{High energy ($>1$keV) spectral energy distribution for \astar\
      and the plausibly associated objects IGR\,J1745.6$-$2901 \citep[from][]{GC:INTEGRAL}
      and HESS\,J1745$-$290 \citep[from][]{HESS:gcprl}.
      Archival X-ray data are shown for the quiescent state: B03 \cite{GC:Baganoff03};
      the largest reported flare seen using Chandra: B01 \cite{GC:Baganoff01}; and the
      largest flux detected using XMM: P03 \cite{GC:Porquet03}. The quiescent state measured
      during these observations is indicated by a `Q'. The \emph{Peak Flare} Chandra spectrum and
      simultaneous H.E.S.S. limit (on a flaring component) are indicated by the flare time of 
      MJD\,53581.94. 
    }	
  \label{fig:2}
\end{figure*}

The simultaneous spectral energy distribution for the Galactic Centre
from these observations is compared, in Fig.~\ref{fig:2}, with 
previous measurements of \astar\ and the possible high energy counterparts
IGR\,J1745.6$-$2901 \citep{GC:INTEGRAL} and HESS\,J1745$-$290.

\section{Conclusion}

The absence of a significant increase in the $>$160~GeV $\gamma$-ray
flux of HESS\,J1745-290 during a major X-ray flare (corresponding to
an increase in flux by a factor of approximately 9 at maximum)
suggests strongly that the keV and TeV emission cannot be attributed
to the same parent population of relativistic particles. A possible
component of the $\gamma$-ray signal that has the same flaring
behaviour as the X-rays is limited to a flux less than 100\% of the
quiescent state signal or $4.2\times10^{-12}$ erg s$^{-1}$ cm$^{-2}$
(2--10 TeV), which should be compared with the $1.9\times10^{-12}$ erg
s$^{-1}$ cm$^{-2}$ 2--10 keV flux during the same period
($\approx$$10\times$ the quiescent flux).  The region of variable
X-ray emission is limited by causality arguments to a size of $r_{X} <
ct_{\rm flare}$ or $\sim$$10^{14}$~cm.  Following
\citet{GC:AtoyanDermer}, the radiation energy density at these
distances from \astar\ is $U_{\rm rad}\,\geq\,3\times10^{-4}$ erg
cm$^{-3}$. If the X-ray emission is interpreted as synchrotron
emission of $\sim$TeV electrons, then the flux limit to inverse
Compton (IC) emission at VHE energies during the flare implies that
$U_{\rm mag} > 0.5\,U_{\rm rad}$ and hence $B>50$ mG in this region.
Since stronger magnetic fields are, in general, expected in this
region \citep[see e.g.][]{GC:Yuan03}, our result does not constrain
models where the X-ray flares are assumed to be produced by
synchrotron emission of relativistic electrons. We note that the
arguments given above assume that the synchrotron flare is caused by
an increase in the number of relativistic electrons. The alternative
explanation that an increase in synchrotron emission occurs due to an
impulsive increase in the magnetic field (with no direct effect on the
IC flux) cannot be excluded.

The fact that HESS\,J1745$-$290 does not appear to be associated with
radiation processes within $10^{14}$~cm (or $\sim$100$\,R_{S}$) of the
supermassive black hole does not exclude all scenarios in which
\astar\ is the acceleration site for the particles responsible for the
TeV emission. Scenarios in which the energy losses of the accelerated
particles occur much farther from \astar\ (for example
\citet{GC:AharonianNeronov05}, \citet{GC:AtoyanDermer} and
\citet{GC:Ballantyne07}) remain viable explanations for this
$\gamma$-ray source.

\begin{acknowledgements}

The support of the Namibian authorities and of the University of
Namibia in facilitating the construction and operation of H.E.S.S. is
gratefully acknowledged, as is the support by the German Ministry for
Education and Research (BMBF), the Max Planck Society, the French
Ministry for Research, the CNRS-IN2P3 and the Astroparticle
Interdisciplinary Programme of the CNRS, the U.K. Science and
Technology Facilities Council (STFC), the IPNP of the Charles
University, the Polish Ministry of Science and Higher Education, the
South African Department of Science and Technology and National
Research Foundation, and by the University of Namibia. We appreciate
the excellent work of the technical support staff in Berlin, Durham,
Hamburg, Heidelberg, Palaiseau, Paris, Saclay, and in Namibia in the
construction and operation of the equipment. We would also like to
thank the anonymous referee for his/her helpful comments.

\end{acknowledgements}

\bibliographystyle{aa} 
\bibliography{general}

\end{document}